\documentclass[conference]{IEEEtran}
\IEEEoverridecommandlockouts
\ifCLASSINFOpdf
\else
\fi
\usepackage{graphicx}
\usepackage{amsmath}
\usepackage{comment}

\begin{document}
%
\title{An Approach for Self-Training Audio Event Detectors Using Web Data}

\author{\IEEEauthorblockN{Benjamin Elizalde\IEEEauthorrefmark{2}\IEEEauthorrefmark{4}, Ankit Shah\IEEEauthorrefmark{1}\IEEEauthorrefmark{4}, Siddharth Dalmia \IEEEauthorrefmark{2}\IEEEauthorrefmark{4}, \\Min Hun Lee\IEEEauthorrefmark{2}\IEEEauthorrefmark{4}, Rohan Badlani\IEEEauthorrefmark{3}\IEEEauthorrefmark{4}, Anurag Kumar\IEEEauthorrefmark{2}\IEEEauthorrefmark{4}, Bhiksha Raj \IEEEauthorrefmark{2} and Ian Lane \IEEEauthorrefmark{2}}
\thanks{\IEEEauthorrefmark{4} First six authors contributed equally.}
\IEEEauthorblockA{Department of Electrical and Computer Engineering, \\
\& Department of Language Technologies Institute,
Carnegie Mellon University, Pittsburgh, PA \IEEEauthorrefmark{2}\\ 
Department of Electronics and Communication, NITK Surathkal, India \IEEEauthorrefmark{1}\\
Department of Computer Science, BITS Pilani, India \IEEEauthorrefmark{3}\\
Email:
\IEEEauthorrefmark{2}bmartin1@andrew.cmu.edu,
\IEEEauthorrefmark{1}ankit.tronix@gmail.com,
\IEEEauthorrefmark{2}sdalmia@andrew.cmu.edu,
\IEEEauthorrefmark{2}mhlee@cmu.edu, \\
\IEEEauthorrefmark{3}rohan.badlani@gmail.com,
\IEEEauthorrefmark{2}alnu@andrew.cmu.edu,
\IEEEauthorrefmark{2}bhiksha@cs.cmu.edu,
\IEEEauthorrefmark{2}lane@cs.cmu.edu}}

\maketitle

\begin{abstract}
Audio Event Detection (AED) aims to recognize sounds within audio and video recordings. AED employs machine learning algorithms commonly trained and tested on annotated datasets. However, available datasets are limited in number of samples and hence it is difficult to model acoustic diversity. Therefore, we propose combining labeled audio from a dataset and unlabeled audio from the web to improve the sound models. The audio event detectors are trained on the labeled audio and ran on the unlabeled audio downloaded from YouTube. Whenever the detectors recognized any of the known sounds with high confidence, the unlabeled audio was use to re-train the detectors. The performance of the re-trained detectors is compared to the one from the original detectors using the annotated test set. Results showed an improvement of the AED, and uncovered challenges of using web audio from videos.

\end{abstract}

%
\IEEEpeerreviewmaketitle

\section{Introduction and Related Work}
\label{sec:Introduction}

Sounds are essential to how humans perceive and interact with the world. Audio content is captured in recordings and shared on the web on a minute-by-minute basis. Academia and industry exploits this acoustic information throughout multiple applications. The dominant application is multimedia video content analysis, where audio is combined with images and text ~\cite{schauble2012multimedia,lew2006content} to index, search and retrieve videos. Another task is human-robot interaction~\cite{maxime2014sound,janvier2012sound}, where sounds complement speech as non-verbal communication. Recently, a growing application is in smart cities~\cite{Salamon:UrbanSound:ACMMM:14}, where sounds are used to detect sources of noise pollution. All of these applications rely on Audio Event Detection (AED) to recognize the occurrence of sounds within audio and video recordings.

The related work on AED has mainly focused on using available datasets to train machine learning models in a supervised manner  ~\cite{stowell2015detection,Salamon:UrbanSound:ACMMM:14,ravanelli_eusipco,piczak2015environmental,DCASE2016workshop}. However, the largest data set, ESC-50~\cite{piczak2015environmental}, contains only 40 samples per class. The numbers strongly contrast with Imagenet, the computer vision counterpart, which has hundreds of samples per class. Hence, training a model which reflects the acoustic diversity of an audio event class is limited. The common solution is to have humans annotating more data. However, the process is costly and slow and thus, other solutions should be explored.

Another solution is to combine the small amount of labeled data with a large amount of unlabeled data. A particular method is semi-supervised self-training, which is an algorithm that iteratively re-trains a model. First, the model is trained using the labeled data set. Then, at each iteration and under a certain criteria, a portion of the unlabeled set could be labeled as any of the known classes. Lastly, using the newly labeled data, the model is re-trained. This approach has been explored for audio events in two papers~\cite{han2016semi,zhang2012semi}. Particularly in~\cite{han2016semi}, the authors collected 17,000 labeled audio-only recordings from \textit{FindSounds.com}. Two thirds were used to train and test a classifier and the rest was treated as unlabeled audio for re-training. The result was an improvement of 1.4\% precision over the baseline, suggesting a valid alternative to improve models. Moreover, the authors pointed out the challenges of utilizing audio-only web recordings. 

In our paper, we followed a similar framework of semi-supervised self-learning, but with the following differences:
\begin{itemize} 
  \item We employed UrbanSounds8k as the labeled set for training and testing. However, we collected YouTube videos as the unlabeled set for re-training, creating mismatch conditions.
  \item For re-training, we used 30 times more audio files.
  \item The unlabeled audio is extracted from videos as opposed to audio-only recordings. Hence, posing challenges during the collection process. For instance, it is not possible to guarantee that the YouTube audio will actually contain any of the sounds.
\end{itemize}  

The paper is structured as follows, in Section~\ref{sec:main} we describe the flow of our self-training approach for sound detectors. Within this Section we describe the sound event dataset and YouTube video collection process and how we pre-processed the audio recordings. Then, we explain how we trained our two machine learning based detectors to compare performance. In Section~\ref{sec:exp} we compare the baseline performance and the self-training performance obtained with different techniques.

\section{Semi-Supervised Self-Training of Audio Event Detectors}
\label{sec:main}
\begin{figure}[h]
   \centering
     \includegraphics[width=0.5\textwidth]{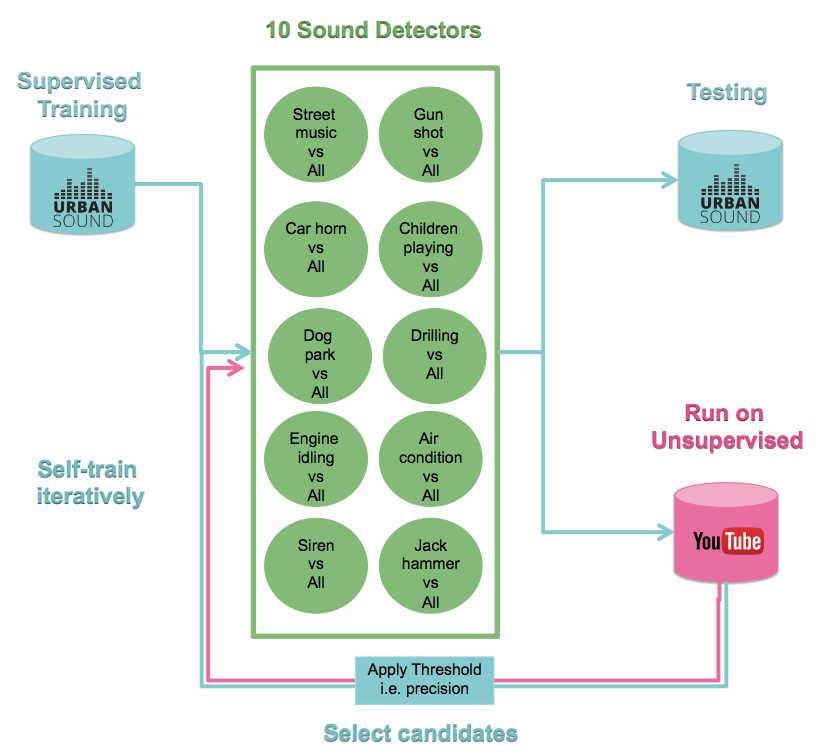}
     \caption{Flow of the semi-supervised self-training of Audio Event Detection. Most of the tuning that improved the performance of the re-trained detectors happened in the selection of candidates.}
     \label{fig:Diagram-Self-learning}
\vspace{-0.1in}
\end{figure}

Semi-supervised self-training is an algorithm that iteratively re-trains a model and our particular framework is illustrated in Figure~\ref{fig:Diagram-Self-learning}. First, using the labeled dataset, the ten class detectors are trained and tested to compute a baseline performance. Second, the unlabeled data is run by the detectors to obtain a class label with its corresponding confidence score. Third, we applied a threshold based on the confidence score to determine candidates for self-training the detectors. Fourth, the detectors are re-trained and again tested on the labeled data to compute the new performance and compare it with the previous. Lastly, steps two, three and four are performed iteratively until the performance converges.

\subsection{Datasets: Labeled and Unlabeled}
\label{Datasets}

\textit{Labeled Data (Training and Testing): UrbanSound8K}\\ 
The UrbanSound8K (US8K) dataset~\cite{Salamon:UrbanSound:ACMMM:14} has 10 classes: \textit{air conditioner, car horn, children playing, dog bark, and street music, gun shot, drilling, engine idling, siren, jackhammer}. The content of the audio may have other overlapping sounds and the target sound may occur in the background or in the foreground. The dataset has about 8,732 audio segments of 3.5 sec average duration. These files are distributed into 10 stratified cross-validation folds. 

\textit{Unlabeled Data (Self-Training): YouTube videos}\\ 
The unlabeled audio comes from YouTube videos and videos are the largest source of audio. The website was chosen because it offers a wide diversity of class samples. The soundtracks of the videos were crawled and downloaded using the Pafy API\footnote{https://pypi.python.org/pypi/pafy}. The audio roughly corresponds to the 10 classes from US8k. The acoustic content is unstructured, and commonly the target sound is occluded by multiple factors such as noise, overlap with other sounds, and channel effects. The web set has 200,000 segments of 3.5 seconds. We converted all of the audio files into raw 16 bit encoding, mono-channel, and 16 kHz sampling rate. 

\textit{Challenges of Unlabeled Data Collection}\\
Audio from videos poses collection challenges. YouTube contains years of videos and in order to process and evaluate audio containing the target sound, the query should serve as a filter. Hence, the query formulation aims to filter in videos roughly matching the ten classes in US8K. Typing a query composed by a noun such as \textit{air conditioner} will not necessarily fetch a video containing such sound event. This happens because the associated tags and metadata are mainly inspired by the video's visual content; contrary to what happens in audio-only websites such as \textit{freesounds.org}. Therefore, we modified the query to be a combination of keywords: ``\textless\textit{audio event label}\textgreater  sound", for example,``air conditioner sound". Although the results empirically improved, another issue was that the audio event was not guaranteed to occur and if present it most likely occurred with a short duration within whole recording. Therefore, we restricted the video length to be larger than five seconds and shorter than ten minutes to reduce the amount of irrelevant audio.

\subsection{Data Preparation}

\textit{Extracting Low-level Features: MFCCs}\\ 
The Mel Frequency Cepstral Coefficients (MFCCs) have been widely used in audio event detection~\cite{metze2014improved,Salamon:UrbanSound:ACMMM:14,piczak2015environmental}. The parameters are standard, such as 10 ms shifts, window of 25 ms and 20 cepstral coefficients including delta and double-delta (time dynamics) for a total of 60 coefficients for each time window or vector.

\begin{figure}[ht]
   \centering
     \includegraphics[width=0.45\textwidth]{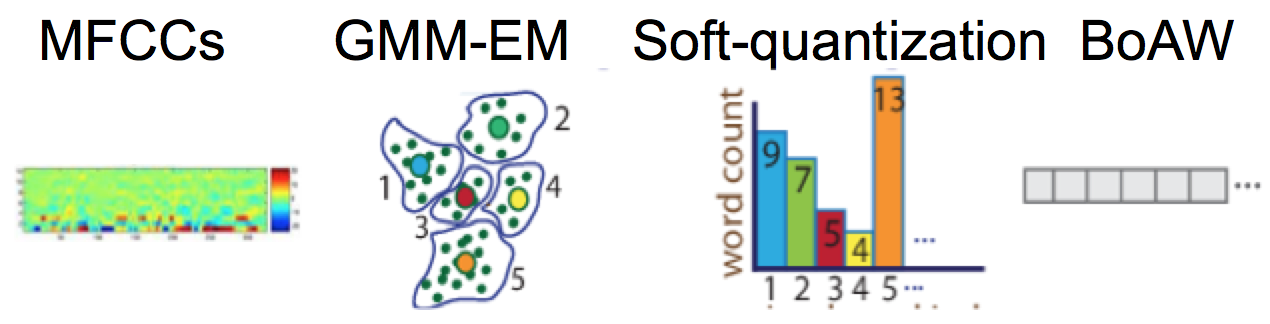}
     \caption{Process of computing our BoW features. The idea behind GMM-based representation is to capture the distribution of MFCC vectors of a recording over the GMM components.}
     \label{fig:Bows}
\vspace{-0.1in}
\end{figure}

\textit{Extracting Intermediate Features: BoAWs}\\ 
An effective approach for characterizing audio events is the Bag-of-Audio-Words (BoAWs) feature representation, which is usually built over low-level features such as MFCCs. The method we followed to compute BoAWs features is broadly illustrated in Figure~\ref{fig:Bows} and detailed in these papers~\cite{bow,elizalde2016experiments}. In the first step, we put all the MFCCs in the training set together. In the second step, we learn an ``audio vocabulary" by grouping the features into ``audio words". In contrast to conventional approaches which uses clustering for grouping words, our method adapts the MFCCs to a Gaussian Mixture Model (GMM) using Expectation Maximization where each mixture represents a word. The third step is quantization, which uses the created vocabulary to turn the MFCC matrix of a given recording into a BoAW histogram-vector of the size of the vocabulary. The conventional quantization process computes the distance of each MFCCs frame to all the audio words and sums the value of one only on the histogram bin that corresponds to the closest word. However, our approach uses soft-quantization, which sums probabilities for all the words.

Formally, let the MFCC vectors of a recording be represented as $\vec{x}_t$. $\vec{x}_t$ is $t^{th}$ $D$ dimensional MFCC frame, $t=1\,\,to\,\,T$. Here, the GMM $\mathcal{G} = \{w_k,N(\vec{\mu}_k, \Sigma_k), k = 1 \,\,to \,\,M\}$, is learned over the MFCCs of training data.  $w_k$, $\vec{\mu}_k$ and $\Sigma_k$ are the mixture weight, mean and co-variance parameters respectively, of the $k^{th}$ Gaussian in $\mathcal{G}$. We train GMM with diagonal co-variance matrices. To obtain the bag of audio word feature representation for any given recording, we first compute the probabilistic assignment to $k^{th}$ Gaussian for each MFCC frame of that recording as in Equation \ref{eq:postr}. This soft assignment is then summed and normalized over all MFCC frames for $k^{th}$ Gaussian as in Eq \ref{eq:histo}. 
\begin{align}
Pr(k | \vec{x}_{t}) & =  \frac{w_{k}N(\vec{x}_{t} ; \vec{\mu}_k, \Sigma_k)}{\sum\limits_{j=1}^M w_jN(\vec{x}_{t} ; \vec{\mu}_k, \Sigma_k)} \label{eq:postr} \\
P(k) & =  \frac{1}{T}\sum\limits_{i=1}^T Pr(k | \vec{x}_{t}) \label{eq:histo}
\end{align}

The final \emph{soft-count} histogram feature representation, represented as $\vec{\alpha}$ is $\vec{\alpha}^M=[P(1),..P(k)..P(M)]^T$. $\vec{\alpha}^M$ features are an $M$-dimensional (M=128) feature representation for any given recording. During testing, the BoAW features are computed in a similar manner however, using the created vocabulary from training.

\subsubsection{Training Detectors: Positive and Negative Classes}
We chose detectors--binary classifiers, because are able to recognize the presence or absence of a particular audio event in a recording. The binary setup also aims to simulate the imbalance ratio of small amount of target sound vs a large amount of non-target sounds, which is common in web retrieval tasks. The audio samples belonging to the target class are referred to as positives and those samples not belonging to the target are referred to as negatives. Each of the ten detectors is trained with both, a positive and a negative class. Positive contains class samples and negative contains samples from the rest of the classes. For instance, the detector for \textit{jackhammer} has all the samples corresponding to \textit{jackhammer} as positives and all the samples corresponding to the other 9 sounds as negatives.

\subsubsection{Training Detectors: SVM} 

One round of experiments was performed with Support Vector Machines (SVMs) because SVMs have been widely explored for sound events~\cite{stowell2015detection}. The ten SVM-based detectors used linear decision boundaries to fit the data and were trained with the intermediate features. To relax the constraints defining the margin of the decision boundary, the parameter ``C" was tuned and set to 0.01. Then, the trained detectors were evaluated using the test set. Although conventional SVMs could employ other techniques to allow non-linear decision boundaries, the problem happens when new audio segments are added for re-training. Each iteration means that the SVM has to be re-trained from scratch using all the train data. The consequence is a bottleneck, which worsens as more segments and iterations are added.

\subsubsection{Training Detectors: NN}

Considering the previous issue, the second round of experiments was performed with Neural Networks (NNs). The NNs are more suitable for the iterative nature of self-training. For example, in order to add a new audio segment for training, the NN does not need to be re-trained from scratch and a quick updating process suffice. The NN-based detectors are also binary classifiers. More precisely, for the NN we utilized a Multi-Layer Perceptron (MLP), with tuned hyper-parameters such as number of layers, neurons, activation function, regularization and loss function. The final architecture consisted of an input of size 128--BoW features dimensionality, one hidden layer of 100 neurons, and two output units-- class or not class. The activation function was ``tanh", the regularization method was dropout (p=0.5), the loss function was cross-entropy and the number of epochs was 10. Then, the trained detectors were evaluated using the labeled test set.

\section{Experiments and Evaluation of Methods}
\label{sec:exp}

\subsection{Computing the Baseline Performance and Running Detectors on Unlabeled Data}
The initial performance computed by our two machine learning algorithms defined the baseline to improve after self-training. The detectors used the labeled data from US8k, which comes divided in $10$ stratified folds. We used $9$ folds as training data and tested on the left-out fold. This is done in $10$ different ways, resulting in $100$ runs for all the 10 events classes and $10$ folds. We evaluated our detectors using average precision as we wanted to detect reliable positive or negative samples. For every class, the average precision (AP) over each fold is computed, as well as the mean AP across all folds referred as Mean AP. Afterwards, the detectors where run on the unlabeled dataset to obtain confidence scores and labels for each of the 200,000 segments. Note that the unlabeled data was carefully handled to be consistent with the 10 fold cross-validation setup. For example, the detectors trained using the first 9 folds may not yield the same performance as the detectors trained with any other fold combination.

\subsection{Selecting Candidates and Self-Training Detectors}
We employed a high confidence threshold to select audio segments as candidates for self-training. The candidates were used for self-training the detectors in combination to the supervised audio segments. Once the detectors were re-trained, they were ran on the supervised test set and their performance was computed. The Mean AP value was compared with the baseline and the whole process was repeated iteratively until the Mean AP converged. 

A key step in the self-training process is the selection of candidates. We tried three main approaches: Detector's output scores, precision and clarity index.

\textit{Score-based} Under this approach, the output of the detector is a probability score that can be interpreted as a confidence value and has been used for self-training in the paper~\cite{han2016semi}. A score threshold of greater or equal than 0.95 was selected to filter in any segment, where 0 means the lowest confidence and 1 means the strongest confidence.

\textit{Precision} 
High precision means that the detector returned more relevant results than irrelevant ones. A precision threshold of greater or equal than 0.95 was set. The value range is the same as the score-based.

\textit{Clarity Index}
Clarity Index (CI), based on the paper~\cite{huang2008active}, aims to determine those segments that are the most confusing for the detector. CI is based on two losses called \emph{relevance loss} and \emph{irrelevance loss}. To understand these losses let us assume that the training data is $\mathcal{D} = \{ (x_1,y_1),(x_2,y_2)...,(x_n,y_n)\}$ and the detector mapping function is denoted by $f$. Let $x^u$ be an unlabeled data point. The Relevance Loss (RL) and the Irrelevance Loss (IL) are defined as 
\begin{align}
RL(x^u,f) = \frac{1}{|\mathcal{D}_0|} \sum_{x_i \in \mathcal{D}_0} I(f(x_i) - f(x^u))\\
IL(x^u,f) = \frac{1}{|\mathcal{D}_1|} \sum_{x_i \in \mathcal{D}_1} I(f(x^u) - f(x_i))
\end{align}
 
The relevance loss is expected to be low if $x^u$ is relevant (positive) and irrelevance loss is expected to be low irrelevant (negative). The difference of the two losses $CI = IL - RL$ is expected to be high (close to 1) for positive instances and low (close to -1) for negative instance. Overall, the CI helps us rank unlabeled segments to choose better segments for self-training. Higher CI implies that $X^u$ is more likely to be positive. An unlabeled point with very high CI would have outscored a large number of training points and hence is expected to be positive. Similarly, lower CI implies the instance is most likely negative.

\section{Results and Discussion} 

\subsection{Baseline} 

The NN outperformed the SVM by an absolute 8.5\% in the baseline performance. The Mean AP score was 57.8\% for SVM and 66.3\% for NN and are shown in Table~\ref{SBER}. One reason for to justify the better performance of the NN, is that it employed nonlinear decision boundaries to fit the data unlike the SVM, which used linear boundaries. As mentioned before, the SVM can also support nonlinear decision boundaries by using kernels, but the computation time was an issue for processing 200,000 segments for the 10 fold combinations, and the re-training.

\subsection{Self-Training}
The main results of this paper are the improvement gain by self-training shown in Table~\ref{SBER} and labeled as SVM Best and NN Best. The overall Mean AP improvement was 1.2\% for both classifiers. Except for SVM's \textit{dog barking}, all the audio events improved their performance. Particularly, \textit{air conditioning} and \textit{jackhammer} benefited the most with about 3\%. More importantly, the performance did not degrade, which is expected to happen when audio that not belonging to the target class is added by re-training. The SVM Best and NN Best results correspond to different threshold types-- CI and Pecision respectively. The performance of the three threshold types was similar (0.5\%-1.4\%) and we cannot say that one should be preferred.

For the three threshold types, tuning affected differently the overall selection of candidates and the detection performance. The number of candidates varied between class from 0 to 2,000 on each iteration. In general, stricter values (greater than 0.9) reduced the number of candidates to two digits. Regarding the detection performance, stricter values (greater than 0.95) and loose values (approximately 0.5) degraded Mean AP, but values close to 0.9 yielded the reported gain. The threshold also defined the number of iterations the algorithm took to converged or stop improving. For our value of 0.9, our algorithm iterated three times. Afterwards, the Mean AP performance converged and then slowly decreased. In general, most of our experiments degraded its performance after several iterations. Two possible explanation are the mismatch conditions and the lack of useful files for self-training.

Mismatch conditions are unavoidable if web audio is intended to be exploited through semi-supervised approaches. There is no control over the recording methods for unlabeled web audio and thus it will most likely be different than the control methods from labeled datasets. In our case, the dataset US8k has different collection methods and acoustic characteristics, which do not match the user-generated YouTube audio. In our experiments, the detectors were self-trained using only the newly labeled ``positive" segments from YouTube. After each iteration, more and more YouTube data was added to the detectors on the positive category but not on the negative. However, the improvement was limited and often degraded. After inspecting some of the rejected files, it seemed that the detectors were discriminating YouTube vs non-YouTube audio, rather than positive vs negative. On the contrary, when ``positive and negative" segments were added, the performance improved.
 
\begin{table}
\begin{center}
{
  \begin{tabular}{ | l | c | c || c | c |}
    \hline
       Category & SVM       & SVM   & NN         & NN    \\
                &  Baseline & Best &  Baseline  & Best\\ \hline\hline
air\_conditioner   &    39.3& 45.1 &49.9&53.2 \\ \hline
car\_horn             & 52.4& 53.0 &51.6&52.8 \\ \hline
children\_playing   &   53.8& 54.3 &65.1&65.2 \\ \hline
dog\_bark      &        76.2& 75.9 &81.7&82.0 \\ \hline
drilling              & 56.6& 57.2 &63.4&63.0 \\ \hline
engine\_idling     &    53.8& 54.1 &68.0&69.8 \\ \hline
gun\_shot        &      67.8& 69.1 &80.4&81.9 \\ \hline
jackhammer    &         60.2& 62.3 &63.7&66.2 \\ \hline
siren           &       72.2& 72.8 &80.2&80.4 \\ \hline
street\_music    &      46.0& 46.4 &58.5&59.0 \\ \hline \hline
Mean AP          &      57.8& 59.0 &66.3&67.5 \\ \hline
 
  \end{tabular} \mbox{} 
}
\end{center}
  \caption{The table shows the class precision and fold average precision (Mean AP). The Mean AP of the SVM and NN baselines was improved through self-training. The SVM Best corresponds to Clarity Index threshold and NN Best corresponds to Precision threshold. }  
\label{SBER}
\end{table} 

\begin{figure}
   \centering
     \includegraphics[width=0.45\textwidth]{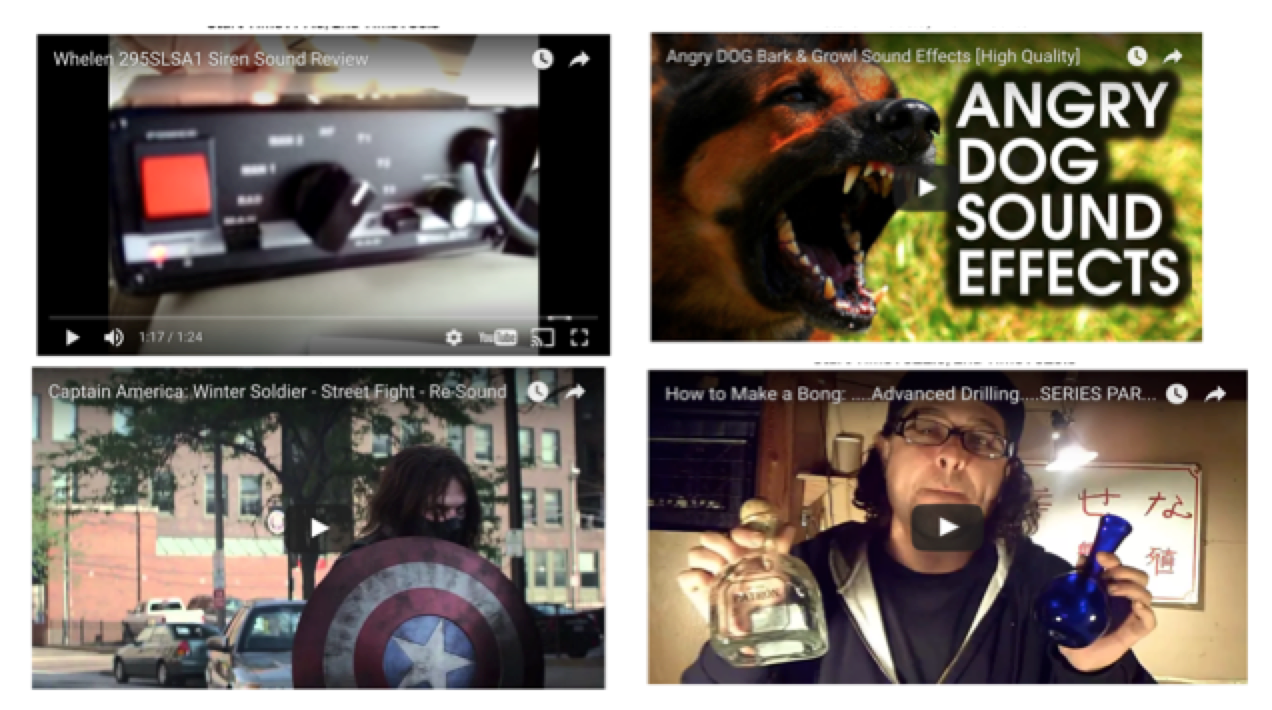}
     \caption{Manual inspection of selected candidates from Siren, Dog bark, Air Conditioner and Drilling.}
     \label{fig:ManualInspection}
\vspace{-0.1in}
\end{figure}

A manual inspection on some of the candidates helped us better understand the audio content used to re-train the detectors. Thumbnails examples are in Figure~\ref{fig:ManualInspection}, illustrating interesting cases. For instance, some videos may have the presence of the sound even though the image didn't corresponded. The first thumbnail-video had the \textit{siren} sound, but the image in the video was just a radio-like box. Another example was when sounds were acoustically similar but semantically different. The third thumbnail-video showed a scene from the movie ``Captain America", where the audio was similar to ``air conditioner", but there was no such item. These examples does not necessarily degrade the quality of the detector as shown in~\cite{elizalde2013audio}. In a similar manner, \textit{gun shot} had, among some of the candidates, object banging sounds. 


\section{Limitations and Future Work}

\textit{Classifier bias} Semi-supervised approaches have limitations related to what extent they can help, as discussed in~\cite{singh2009unlabeled}. Especially, self-training has an inherent detector bias issue which happens when a detector is trained with an initial set of data. The detector then, is ran on the unlabeled data and the confidence score depends on the initial model. Once we add new segments, we are enforcing the acoustic characteristics of the previous model and not necessarily making our models more robust. Addressing the issue was out of scope, but could be a reason for the fast convergence in our results.

\textit{Threshold type} The set of thresholds utilized are a reasonable approach supported in the literature. However, a more elaborated objective function should be considered to better select candidates.

\section{Conclusions} 
In this work we proposed a framework of semi-supervised self-training of audio event detectors, where the detectors were trained with the annotated US8K dataset, and the self-training employed unlabeled audio from YouTube videos. The NN detectors yielded a higher baseline performance than SVM. Both detectors and almost all the classes benefited from self-training. Despite the audio mismatch conditions and the possibility of having few or no target sounds to be candidates, the performance after self-training did not degrade. Further exploration to select candidates offers a valuable opportunity. Unlabeled audio from videos can help audio event detection.




%
 
\bibliographystyle{IEEEtran}
\bibliography{EUSIPCO}

\begin{thebibliography}{10}
\providecommand{\url}[1]{#1}
\csname url@samestyle\endcsname
\providecommand{\newblock}{\relax}
\providecommand{\bibinfo}[2]{#2}
\providecommand{\BIBentrySTDinterwordspacing}{\spaceskip=0pt\relax}
\providecommand{\BIBentryALTinterwordstretchfactor}{4}
\providecommand{\BIBentryALTinterwordspacing}{\spaceskip=\fontdimen2\font plus
\BIBentryALTinterwordstretchfactor\fontdimen3\font minus
  \fontdimen4\font\relax}
\providecommand{\BIBforeignlanguage}[2]{{%
\expandafter\ifx\csname l@#1\endcsname\relax
\typeout{** WARNING: IEEEtran.bst: No hyphenation pattern has been}%
\typeout{** loaded for the language `#1'. Using the pattern for}%
\typeout{** the default language instead.}%
\else
\language=\csname l@#1\endcsname
\fi
#2}}
\providecommand{\BIBdecl}{\relax}
\BIBdecl

\bibitem{schauble2012multimedia}
P.~Sch{\"a}uble, \emph{Multimedia information retrieval: content-based
  information retrieval from large text and audio databases}.\hskip 1em plus
  0.5em minus 0.4em\relax Springer Science \& Business Media, 2012, vol. 397.

\bibitem{lew2006content}
M.~S. Lew, N.~Sebe, C.~Djeraba, and R.~Jain, ``Content-based multimedia
  information retrieval: State of the art and challenges,'' \emph{ACM
  Transactions on Multimedia Computing, Communications, and Applications
  (TOMM)}, vol.~2, no.~1, pp. 1--19, 2006.

\bibitem{maxime2014sound}
J.~Maxime, X.~Alameda-Pineda, L.~Girin, and R.~Horaud, ``Sound representation
  and classification benchmark for domestic robots,'' in \emph{2014 IEEE
  International Conference on Robotics and Automation (ICRA)}.\hskip 1em plus
  0.5em minus 0.4em\relax IEEE, 2014, pp. 6285--6292.

\bibitem{janvier2012sound}
M.~Janvier, X.~Alameda-Pineda, L.~Girinz, and R.~Horaud, ``Sound-event
  recognition with a companion humanoid,'' in \emph{2012 12th IEEE-RAS
  International Conference on Humanoid Robots (Humanoids 2012)}.\hskip 1em plus
  0.5em minus 0.4em\relax IEEE, 2012, pp. 104--111.

\bibitem{Salamon:UrbanSound:ACMMM:14}
J.~Salamon, C.~Jacoby, and J.~P. Bello, ``A dataset and taxonomy for urban
  sound research,'' in \emph{22st {ACM} International Conference on Multimedia
  ({ACM-MM'14})}, Orlando, FL, USA, Nov. 2014.

\bibitem{stowell2015detection}
D.~Stowell, D.~Giannoulis, E.~Benetos, M.~Lagrange, and M.~D. Plumbley,
  ``Detection and classification of acoustic scenes and events,'' \emph{IEEE
  Transactions on Multimedia}, vol.~17, no.~10, pp. 1733--1746, 2015.

\bibitem{ravanelli_eusipco}
M.~Ravanelli, B.~Elizalde, K.~Ni, and G.~Friedland, ``Audio concept
  classification with hierarchical deep neural networks,'' in \emph{Proceedings
  of {EUSIPCO}}, 2014.

\bibitem{piczak2015environmental}
K.~J. Piczak, ``Environmental sound classification with convolutional neural
  networks,'' in \emph{2015 IEEE 25th International Workshop on Machine
  Learning for Signal Processing (MLSP)}.\hskip 1em plus 0.5em minus
  0.4em\relax IEEE, 2015, pp. 1--6.

\bibitem{DCASE2016workshop}
T.~Virtanen, A.~Mesaros, T.~Heittola, M.~Plumbley, P.~Foster, E.~Benetos, and
  M.~Lagrange, \emph{Proceedings of the Detection and Classification of
  Acoustic Scenes and Events 2016 Workshop (DCASE2016)}.\hskip 1em plus 0.5em
  minus 0.4em\relax Tampere University of Technology. Department of Signal
  Processing, 2016.

\bibitem{han2016semi}
W.~Han, E.~Coutinho, H.~Ruan, H.~Li, B.~Schuller, X.~Yu, and X.~Zhu,
  ``Semi-supervised active learning for sound classification in hybrid learning
  environments,'' \emph{PloS one}, vol.~11, no.~9, p. e0162075, 2016.

\bibitem{zhang2012semi}
Z.~Zhang and B.~Schuller, ``Semi-supervised learning helps in sound event
  classification,'' in \emph{Acoustics, Speech and Signal Processing (ICASSP),
  2012 IEEE International Conference on}.\hskip 1em plus 0.5em minus
  0.4em\relax IEEE, 2012, pp. 333--336.

\bibitem{metze2014improved}
F.~Metze, S.~Rawat, and Y.~Wang, ``Improved audio features for large-scale
  multimedia event detection,'' in \emph{Multimedia and Expo (ICME), 2014 IEEE
  International Conference on}.\hskip 1em plus 0.5em minus 0.4em\relax IEEE,
  2014, pp. 1--6.

\bibitem{bow}
F.-F. Li and P.~Perona, ``The perceived position of moving objects:
  Transcranial magnetic stimulation of area {MT+} reduces the flash-lag
  effect,'' in \emph{IEEE CVPR}, vol.~2, 2005.

\bibitem{elizalde2016experiments}
B.~Elizalde, A.~Kumar, A.~Shah, R.~Badlani, E.~Vincent, B.~Raj, and I.~Lane,
  ``Experiments on the dcase challenge 2016: Acoustic scene classification and
  sound event detection in real life recording,'' in \emph{DCASE2016 Workshop
  on Detection and Classification of Acoustic Scenes and Events}, 2016.

\bibitem{huang2008active}
T.~S. Huang, C.~K. Dagli, S.~Rajaram, E.~Y. Chang, M.~I. Mandel, G.~E. Poliner,
  and D.~P. Ellis, ``Active learning for interactive multimedia retrieval,''
  \emph{Proceedings of the IEEE}, vol.~96, no.~4, pp. 648--667, 2008.

\bibitem{elizalde2013audio}
B.~Elizalde, M.~Ravanelli, and G.~Friedland, ``Audio concept ranking for video
  event detection on user-generated content,'' \emph{in Multimedia (SLAM
  2013)}.

\bibitem{singh2009unlabeled}
A.~Singh, R.~Nowak, and X.~Zhu, ``Unlabeled data: Now it helps, now it
  doesn't,'' in \emph{Advances in neural information processing systems}, 2009,
  pp. 1513--1520.

\end{thebibliography}

\end{document}